\documentstyle[sprocl]{article}
\def\be{\begin{equation}}
\def\ee{\end{equation}}

\begin{document}

\title{ENHANCED CORRELATIONS AND WIDE CHARGE DISTRIBUTIONS IN PION PRODUCTION}

\author{I.V.ANDREEV}

\address{ P.N.Lebedev Physical Institute,Leninsky pr.53,117924 Moscow,Russia}

\maketitle\abstracts{
Unusual phenomena in pion production are considered.It is argued that pions
may be in a squeezed state having enhanced identical pion correlations.
Charge distribution in soft chiral pion bremsstrahlung is shown to be
very broad.}

\section{Introduction}

High-energy interactions are dominated by multiple production processes.
It is of interest to consider possibilities of unusual phenomena in multiple
pion production which may shed light on the nature of the process of
interaction.Here we consider two such phenomena,concerning identical pion
correlations and charge distributions.

\section{Enhanced correlations in chaotic squeezed states}

Bose-Einstein correlations (BEC) of statistically produced particles
are of great interest in particle physics because their measurement
gives a possibility to recover space-time region of interaction
(the particle source size).
According to common belief,the maximal rise of tho-particle distributions
of identical particles is equal to two (BEC effect).However larger
correlations
are not excluded.Below it will be demonstrated that this is the case for
correlations in chaotic squeezed states~\cite{AW96}.

 First we present a simple model of squeezed state production.
Consider decay of pionic matter arising in the course of particle collision.
Pions in matter (quasiparticles) are described by creation-annihilation
operators $b^{\dag},b$ and have a spectrum $E_b$.Free pions at space infinity
are described by $a^{\dag},a$ operators and have the usual spectrum
$E_a=\sqrt{p^2+m^2_{\pi}}$.Redecomposing the final state of pionic matter
in terms of free pions,we have:
\be
a=b\cosh{r}+b^{\dag}\sinh{r} , a^{\dag}=b\sinh{r}+b^{\dag}\cosh{r}
\label{eq:sq}
\ee
with
\be
 r=\frac{1}{2}\log(E_a/E_b)
\label{eq:rl}
\ee
This is a squeezing transformation with the squeezing parameter $r(p)$.
If the initial quasipions are in coherent state,then the final pions
are in squeezed state.

 We suggest that pionic matter is chaotic (for example it may be thermal)
So it is represented as a random set of coherent (or squeezed) states.
The averaging of operators is performed with a Gaussian weight,see
 Ref.~\cite{APW93}.Here the blob of decaying pionic matter is
characterized by two quantities entering the weight:

 (a)The function
$f({\bf x})$ describing the space form of the pion source (we take here
stationary case for simplicity).It has a characteristic size R (which is
of the main interest in BEC study) and its Fourier transform
$f({\bf k})$ represents corresponding form-factor.

 (b)Primordial correlator
of pionic matter which has its own scale and which reduces to number density
$n({\bf k})$ for stationary source.

 Now evaluation of averaged matrix elements is straightforward.Single-particle
inclusive cross-section is:
\begin{eqnarray}
\rho_{1}({\bf p})=\frac{(2\pi)^{3}}{\sigma}\frac{d\sigma}{d^{3}p}=
<a^{\dag}({\bf p})a({\bf p})>\nonumber\\
=\int\frac{d^{3}k}{(2\pi)^{3}}[n_{b}({\bf k})\cosh{2r({\bf k})}+
\sinh^{2}r({\bf k})]f({\bf {p-k}})f({\bf {k-p}})
\label{eq:rho}
\end{eqnarray}
showing enhancement arising due to squeezing.Our main result here is
two-particle inclusive distribution:
\be
    \rho_{2}({\bf p}_{1},{\bf p}_{2})=
\rho_{1}({\bf p}_{1})\rho_{1}({\bf p}_{2})+
\vert<a^{\dag}({\bf p}_{1})a({\bf p}_{2})>\vert^{2}+
\vert<a({\bf p}_{1})a({\bf p}_{2})>\vert^{2}
\label{eq:rho2}
\ee
with
\be
<a^{\dag}({\bf p}_{1})a({\bf p}_{2})>=
\int\frac{d^{3}k}{(2\pi)^{3}}[n_{b}({\bf k})\cosh{2r({\bf k})}+
\sinh^{2}r({\bf k})]f({\bf p_{1}-k})f({\bf k-p_{2}}),
\label{eq:adaga}
\ee
\be
<a({\bf p_{1}})a({\bf p_{2}})>=
\int\frac{d^{3}k}{(2\pi)^{3}}[n_{b}({\bf k})+\frac{1}{2}]
\sinh{2r({\bf k})}f({\bf k-p_{1}})f({\bf k-p_{2}})
\label{eq:aa}
\ee
The third term in Eq.~\ref{eq:rho2} represents an additional enhancement
arising due to squeezing.It is easy to see that relative two-particle
inclusive distribution
\be
C_{2}({\bf p_{1}},{\bf p_{2}})=\rho_{2}({\bf p_{1}},{\bf p_{2}})/
\rho_{1}({\bf p_{1}})\rho_{1}({\bf p_{2}})
\label{eq:c2}
\ee
may be any in this case,there is no necessity for $C_{2}$ to be
less than two.

\section{Very broad $\pi^{0}$ and $\pi^{ch}$ distributions}

Charge distributions of pions in multiple production processes drew much
attention recently.This is due to expectations to detect disoriented chiral
condensate (DCC) formation in high energy collisions~\cite{AR91,B92}.
The coherent pulses of semiclassical pion field are suggested to be emitted
leading to anomalously large fluctuations in the ratio of neutral to charge
pions produced.In particular the probability to produce $n_{0}$ neutral
pions is given by inverse square root formula,
\be
w(n_{0})\sim 1/\sqrt{n_{0}n}
\label{eq:wn}
\ee
being very flat and so quite different from usual binomial-like distributions.

Now the problem arises--to what extent the Eq.~\ref{eq:wn} may be considered
as a signature of DCC formation.Let us remind in this connection that
the distribution of Eq.~\ref{eq:wn} was found long ago in a model of
independent coherent
pion production with isotopic spin conservation taken into account~\cite{A81}.
Here we consider soft chiral pion bremsstrahlung accompanying some basic
high energy process and estimate charge distribution of the pions.
It will be found that neutral pion number distribution again has the form
of Eq.~\ref{eq:wn}

The soft chiral pion bremsstrahlung was studied many years ago~\cite{WB70}.
Similarly to photons,soft pions are emitted from external lines of diagrams
representing the basic process.The net result for soft pion emission
is given by substitution~\cite{WB70}:
\be
\psi_{j} \to \exp(-i{\bf\pi}{\bf X})\psi_{j}
\label{eq:psi}
\ee
for every incoming and outgoing line,where $\pi_{i}$ is the pion field,
$X_{i}=\gamma_{5}\tau_{i}/2f_{\pi}$ for fermions,$f_{\pi}=93MeV$
is the pion decay constant.

Here we are interested in charge distribution of the pions and calculate
matrix elements
\be
M=<n_{+},n_{-},n_{0}\vert\exp(-i{\bf\pi}{\bf X})\Gamma\exp(i{\bf\pi}{\bf X})
\vert 0>
\label{eq:ma}
\ee
for simple two-particle vertices $\Gamma$.The momentum transfer $\Delta p$
in the vertex $\Gamma$ is suggested to be large.

The scalar $(\Gamma=I)$ and electromagnetic $(\Gamma=(\tau_{3}+N_{B})/2)$
vertices have been considered~ \cite{ABNS96}.In both cases the total pion
distribution was found to be narrow (Poisson-like) whereas the neutral
(and charged) pion number distributions $w(n_{0})$ appeared to be very
flat.In both cases they reproduce inverse square root behaviour
of Eq.~\ref{eq:wn} ,thus indicating that this distribution is typical for
coherent soft pions and do not indicate directly on DCC formation.

So we conclude that one has to look for more delicate criteria of DCC
formation.Here the energy distribution of the pions produced may appear
helpful.In the case of pion bremsstrahlung it rises sharply with energy,
$dE\sim k^{2}dk$,and high momentum transfers $\Delta p$ (and so the
presence of high $p_{T}$ particles in high-energy events) are highly
favourable for copious production of the soft pions.

\section*{Acknowledgements}
This work was supported by Russian Fund for Fundamental Research
(grant 96-02-16210a).

\end{document}